# First-Principles Studies of the Atomic, Electronic, and Magnetic Structure of α-MnO$_2$ (Cryptomelane)

Eric Cockayne[*] and Lan Li
*Ceramics Division, Material Measurement Laboratory,
National Institute of Standards and Technology, Gaithersburg, Maryland 20899 USA*
*Corresponding author. Tel: (+1) 301-975-4347; Fax: (+1) 301-975-5334; electronic address: eric.cockayne@nist.gov

**Abstract**

Density functional theory calculations are used to investigate α-MnO$_2$, a structure containing a framework of corner and edge sharing MnO$_6$ octahedra with tunnels in between. Placing K$^+$ ions into the tunnels stabilizes α-MnO$_2$ with respect to the rutile-structure β-MnO$_2$ phase, in agreement with experiment. The computed magnetic structure has antiferromagnetic (ferromagnetic) Mn-Mn interactions between corner-sharing (edge-sharing) octahedra. Pure α-MnO$_2$ is found to be a semiconductor with an indirect band gap of 1.3 eV. Water and related hydrides (OH$^-$; H$_3$O$^+$) can also be accommodated in the tunnels; the equilibrium K-O distance increases with increasing oxygen hydride charge.

**1. Introduction**

Manganese dioxide (MnO$_2$)-based materials are of great interest for various applications, ranging from catalysts and batteries to energy efficient devices and carbon storage applications [1-6]. Mn is multivalent, and thus forms oxides of several different stoichiometries [7]. For MnO$_2$ phases, the oxidation state is Mn$^{4+}$. The equilibrium phase of MnO$_2$ at standard temperature and pressure is β-MnO$_2$ [7], or pyrolusite [8], with the rutile structure, but several metastable phases are also known.

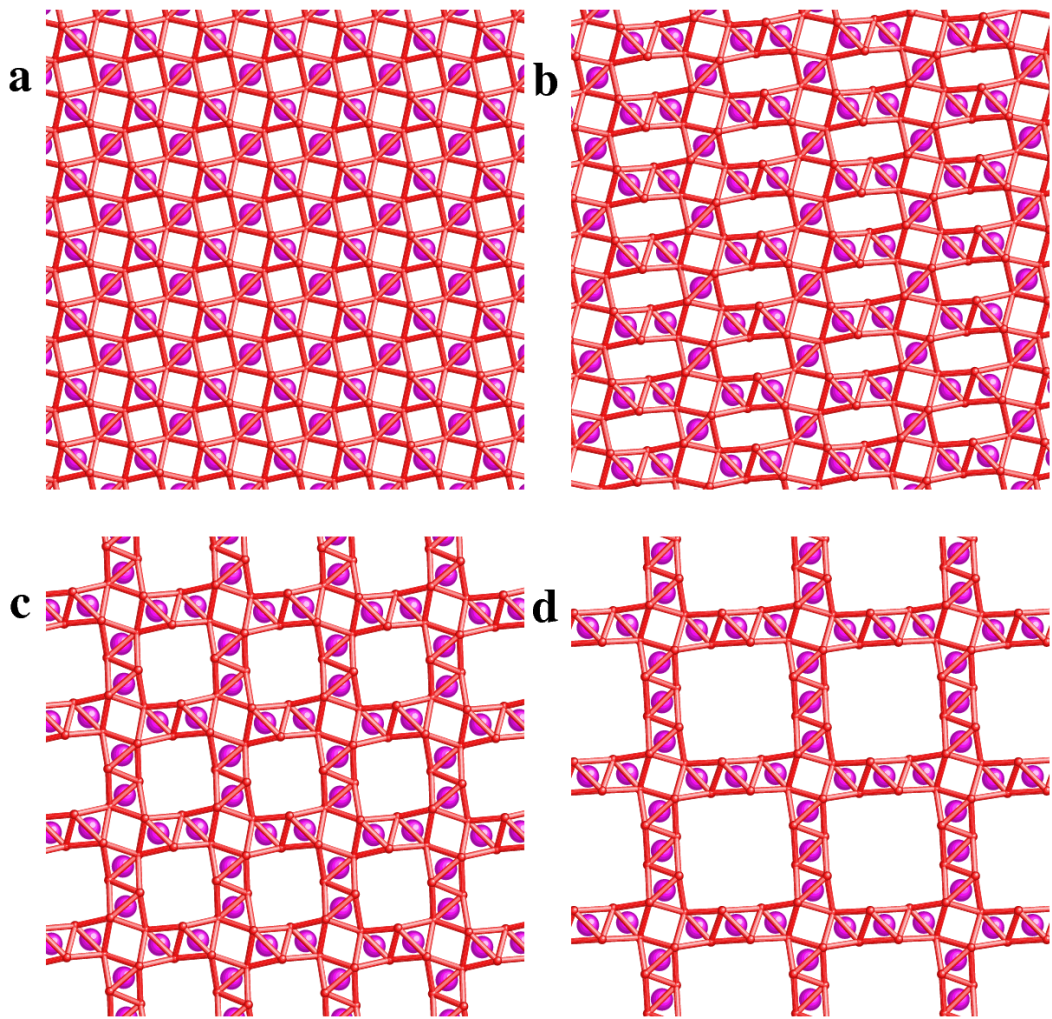

Fig. 1 Octahedral molecular sieve $MnO_2$ structures with alternating 1 x 1 and $m$ x $n$ tunnels. (a) $m=1$, $n=1$ β-$MnO_2$ structure. (b) $m = 2$, $n=1$. (c) $m=2$, $n=2$ α-$MnO_2$ structure. (d) $m=3$, $n=3$. Mn atoms shown in purple and oxygen octahedral frameworks in red.

In fact, $MnO_2$ forms the framework of an entire family of "octahedral molecular sieve" (OMS) structures (Fig. 1) [4,9]. The building blocks of these structures are columns of edge-sharing $MnO_6$ octahedra. These columns join either corner-to-corner or edge-to-edge. As shown in Fig. 1, $MnO_2$ can form an infinite number of OMS structures with alternating (1 x 1) tunnels and ($m$ x $n$) tunnels. The $m=1$, $n=1$ OMS structure is β-$MnO_2$. The $m=2$, $n=2$ OMS structure is the well-known α-$MnO_2$ phase [8]. Careful studies of α-$MnO_2$ show that the structure generally contains additional species, such as cations ($K^+$, $Pb^{2+}$, $Ba^{2+}$, etc.), or water molecules [8-10] inside the 2x2 tunnels. The presence of these additional species offers an opportunity to design $MnO_2$ OMS materials with tailored properties. Depending on which species are present, α-$MnO_2$ is known by various names such as hollandite or cryptomelane [8-9]. In this work, we focus on the $K^+$ containing, or cryptomelane variant.

To fully exploit the properties of MnO$_2$ OMS materials, a fundamental understanding of their atomic and electronic structure is needed. In recent years, there have been considerable experimental studies using Raman scattering spectroscopy, X-ray diffraction, and X-ray photoelectron spectroscopy [11-13]. Theoretical studies can complement these experimental observations. In particular, first-principles density functional theory (DFT) calculations can uncover the electronic origin of structure-property relationships in these advanced materials. In this work, we use DFT calculations to investigate the structure and energetics of MnO$_2$ OMS materials, in particular α-MnO$_2$.

The paper is organized as follows. In Sec. 2, we present the computational methods and discuss the treatment of Mn magnetism within DFT. In Sec. 3 and 4, we present our results and conclusions on the atomic, electronic, and magnetic structure of α-MnO$_2$, with and without additional species present that, for simplicity, we term "dopants" in this work.

## 2. Computational Methods

Structural and electronic structure calculations were performed using the Vienna *Ab initio* Simulation Package (VASP) code [14] based on self-consistent density functional theory (DFT). We used projector-augmented wave pseudopotentials [15] in conjunction with a plane wave expansion of the wavefunctions. The generalized gradient approximation (GGA) was used to approximate the exchange and correlation functional, using the recently-developed PBEsol (Perdew-Burke-Ernzerhof revised for solids [16]) parameterization.

We utilized convergence tests to select the *k*-point mesh size and plane-wave energy cutoff. Structural geometries and forces were well-converged for a 2×6×2 Monkhorst-Pack grid and a 400 eV cutoff. An 8x24x8 Monkhorst-Pack mesh was used for density of state (DOS) calculations. A Gaussian smearing of 0.05 eV was used for the Fermi surface broadening. Relaxations of atomic positions and lattice vectors were performed until residual forces were 0.01 eV/Å or less.

Because Mn$^{4+}$ is a magnetic ion, it is crucial to include the effects of magnetism in electronic structure studies of manganese oxides. Previous works have shown that the exchange-correlational approach used within DFT has a large effect on the computed electronic structure and the magnetic ordering of manganese oxides [17]. These studies predated the PBEsol version of GGA. Using PBEsol for exchange and correlation, we first revisited β-MnO$_2$ within the GGA+U(+J) approach for magnetism. The experimentally-known magnetic structure for β-MnO$_2$ is straightforward: the Mn-Mn interaction for corner-sharing MnO$_6$ octahedra is antiferromagnetic [18] due to superexchange. This completely determines the magnetic structure, as shown in Fig. 2(a). Remarkably, using the GGA+U+J approach of Liechtenstein *et al*. [19], we were able to reproduce both the band gap = 0.27 eV [20] and the unit cell volume [10] simultaneously by using reasonable values for the effective Mn on-site Coulomb (U) and exchange (J) values of 2.8 eV and 1.2 eV, respectively. These parameters are assumed to be transferrable to $Mn^{4+}$ ions in other geometries such as α-$MnO_2$.

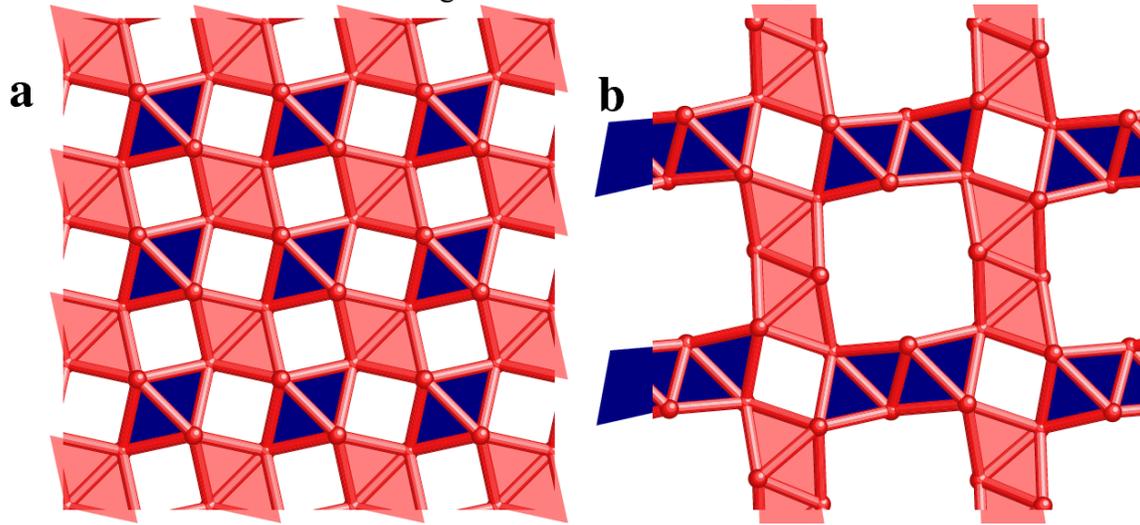

Fig. 2 (a) Experimental magnetic structure of β-$MnO_2$. (b) Predicted ground state magnetic structure of α-$MnO_2$. Octahedra are colored according to the spin state of the central Mn, dark blue for spin "up" and light red for spin "down". All octahedra in each column have the same spin.

The density of states (DOS) and local density of states (LDOS) show the electronic structure of α-$MnO_2$ and how it is affected by dopants. In VASP, the LDOS is obtained via projecting the DOS in spheres of Wigner radii centered at nuclei. Both are spin-dependent in our calculations. The total DOS is the sum of the spin-up and spin-down DOS.

## 3. Results and Discussion

3.1. Magnetism

Because of the coupling between magnetic ordering, energetics, and structure, it is first necessary to determine a model for the magnetic ordering in α-$MnO_2$. We used the GGA+U+J approach discussed above and applied it to several models for the magnetic ordering. For simplicity, we only considered colinear magnetism. The lowest-energy magnetic state found is shown in Fig. 2(b). Not surprisingly, the Mn-Mn coupling between corner-sharing $MnO_6$ is antiferromagnetic (AFM). The interactions between edge-sharing octahedra in neighboring columns are found to be weakly ferromagnetic. Assuming that the calculated magnetic structure is accurate, thermal fluctuations above cryogenic temperatures would readily randomize the magnetic interactions between edge-sharing columns. While the antiferromagnetic interactions are much stronger, they are constrained in quasi-one-dimensional units formed by the four columns of $MnO_6$ octahedra that surround a 1x1 tunnel. In analogy with the 1D Ising model, the AFM ordering at room temperature is expected to be only short-range.

3.2. Structure

α-MnO$_2$ is found to have either a tetragonal structure or a related monoclinic structure with pseudotetragonal symmetry [9]. Gao et al. [11] reported a tetragonal structure at composition K$_{0.11}$MnO$_2$[H$_2$O,H$_3$O]$_{0.07}$ with $a$ = 9.8241 and $c$ = 2.856 (all unit cell parameters in this work in Å), whereas a work related to the present study [6] reports a monoclinic cell at composition K$_{0.09}$MnO$_2$(H$_2$O)$_{0.08}$ with $a$ = 9.8394, $b$ = 2.856, $c$ = 9.790, and β = 90.138º. The different crystal symmetries may be related to the different "dopant" concentrations reported in the two studies. Our minimum-energy calculated structure for pure α-MnO$_2$ structure is monoclinic with $a$ = 9.702, $b$ = 2.856, $c$ = 9.685, β = 90.041º (Fig. 3). The Mn-O bond lengths are all ~ 1.90 Å. The calculated energy difference between the monoclinic and parent tetragonal structures is only 0.1 meV per MnO$_2$ formula unit. Although the calculated structure has a smaller unit cell than the experiment, these calculations are performed without dopants; as shown below, much better agreement is obtained when the dopants are included. In the above-cited monoclinc refinement [6], there are 0.72 K and 0.634 H$_2$O per unit unit cell of composition Mn$_8$O$_{16}$. To mimic the partial occupancy of the dopants with an explicit atomistic model, we use a α-MnO$_2$ cell tripled along the $b$ axis. The experimental stoichiometry is then closely matched by placing 2 K and 2 H$_2$O within the tripled cell. Before optimizing the arrangement of K and H$_2$O, we first explore the energetics and electronic structure of α-MnO$_2$ with K doping alone.

3.3 K dopants

As determined by structure refinements, K$^+$ ions in α-MnO$_2$ fit in the 2x2 tunnels. We find the lowest-energy position of a K$^+$ ion to be at the experimental position with eight oxygen near neighbors (Fig. 4 (a)) [6,9,11]. The distances between K$^+$ and its closest Mn and O are 3.57 Å and 2.85 Å respectively. The energy of a K$^+$ halfway between two neighboring equilibrium positions is 0.37 eV higher. We use this result as an estimate for the energy barrier for K$^+$ diffusion in α-MnO$_2$. Because the barrier to K diffusion is small, we propose that during the synthesis of α-MnO$_2$, individual K atoms might be adsorbed to the open end of (2 × 2) tunnels and then diffuse throughout the tunnels.

Formally, the addition of a K$^+$ to MnO$_2$ reduces one Mn$^{4+}$ to Mn$^{3+}$. Our calculations show that the electron donated by the K is shared by several Mn leading to noninteger charges on these Mn. We investigated possible electronic structures where the extra electron is initially located on a single Mn, and the environment around this Mn given a Jahn-Teller distortion typical for Mn$^{3+}$ ions [21]. With the ions frozen in this position, the state with a localized electron is found to be electronically stable. Once lattice relaxation is allowed, however, the magnetic structure reverts back to what was found initially. Whether this result is an artifact of the treatment of magnetism in DFT or whether donated electrons in α-MnO$_2$ are indeed shared by more than one Mn remains to be resolved.

We next investigated the binding energy of K. Neglecting finite temperature effects, the K-MnO$_2$ binding energy is computed as

$$E_b = (E_{\text{total, doped}} - E_{\text{total, undoped}} - nE_K) / n \quad (1)$$

where the first two terms refer to the total energies of Mn$_{24}$O$_{48}$ with and without K doping, respectively; n specifies the number of K; and $E_K$ is the energy of an individual K atom. The computed binding energy of 4.36 eV implies a strong interaction between K and α-MnO$_2$.

A related question is the comparative stablility of α-MnO$_2$ and β-MnO$_2$. We find that β-MnO$_2$ is more stable by about 20 meV per MnO$_2$ formula unit, in agreement with the experimental phase diagram. But when K is added, there is no favorable interstitial position in β-MnO$_2$. The optimal position found is within the 1x1 tunnels (Fig. 4(b)), but that leads to large structural distortions. The distances between K and its closest Mn and O are 2.78 Å and 2.30 Å respectively, much shorter than those in α-MnO$_2$. For composition K$_{1/24}$MnO$_2$, the structure based on the α-phase is 342 meV lower in energy per formula unit than that based on the β-phase. Our results imply a crossover in stability for K$_x$MnO$_2$ at very low K concentration $x$ = 0.002. These results are fully consistent the experimental observations that α-MnO$_2$ requires dopants for stability.

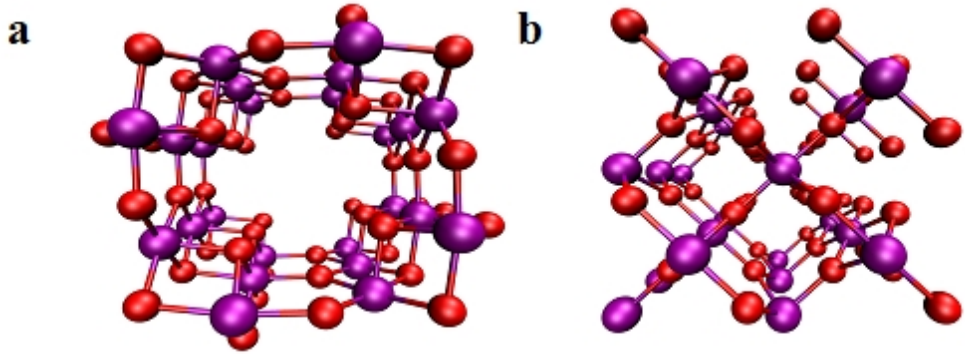

Fig. 3. Crystal structures of (a) α-MnO$_2$ and (b) β-MnO$_2$, Mn in purple and O in red.

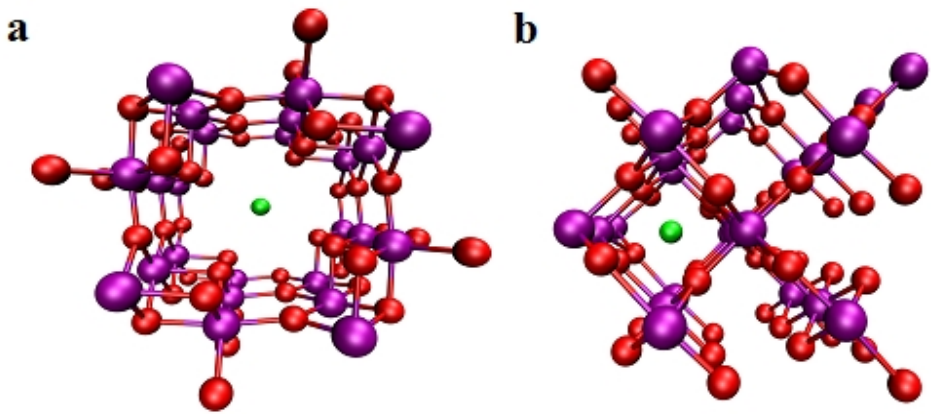

Fig. 4. K-doped α- and β-MnO$_2$, Mn in purple, O in red and K in green: (a) a cell of K-doped α-Mn$_{24}$O$_{48}$ and (b) a cell of K-doped β-Mn$_{24}$O$_{48}$.

To further understand the effect of K on α-MnO$_2$, we study the changes in electronic structure when K is doped into α-MnO$_2$. Fig. 5 shows the total DOS of undoped α-MnO$_2$ as well as the LDOS of α-MnO$_2$ after doping with a K atom. The Fermi level ($E_f$) is shifted to zero in each case. Undoped α-MnO$_2$ is a semiconductor with the Fermi level inside the band gap. The band gap is 1.33 eV (Fig. 3a). Detailed band structure computations (not shown) show that the highest valence band state is at Γ (the Brillouin zone center) and the lowest conduction band state is a zone-edge state at **k** = (0.5,0,0.5) in reciprocal lattice units. With K doping, the conduction band of α-MnO$_2$ is partially filled (Fig. 5(b)). This partial filling of the conduction band is a consequence of electronic charge transfer from K to α-MnO$_2$. Integrating the LDOS of doped α-MnO$_2$ up to the Fermi energy, the estimated charge transferred toward α-MnO$_2$ is 1.22 electrons. The band gap decreases slightly to 1.29 eV (for composition K$_{1/6}$MnO$_2$). Experimentally, the band gap for cryptomelane is 1.32 eV [11], in excellent agreement with our calculations, demonstrating their predictive power.

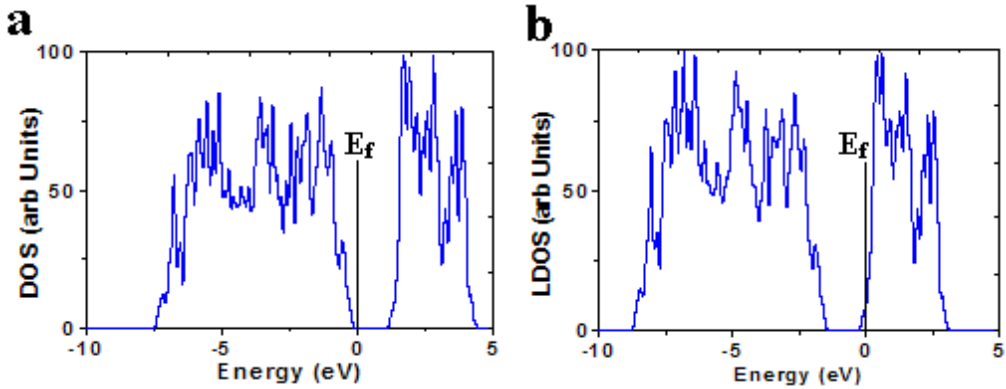

Fig. 5 Electronic structure calculations of undoped and K-doped α-MnO$_2$, Fermi level $E_f$ = 0: (a) Total DOS of undoped α-MnO$_2$; and (b) DOS of α-MnO$_2$ after doping with a K atom.

3.4 Interactions of K-doped α-MnO$_2$ with OH$^-$, H$_2$O and H$_3$O$^+$

According to experimental observations, the synthesis of K-doped α-MnO$_2$ materials may also incorporate other species such as H$_2$O in the 2 x 2 tunnels. Since x-ray diffraction can not resolve H positions, there is some question about whether O inside the tunnel occurs as a neutral water molecule or in the form of another hydride such as OH$^-$ or H$_3$O$^+$. Gao et al. [11] suggest that H$_3$O$^+$ may be present, possibly because positive cations such as K$^+$ tend to occupy the tunnels. On the other hand, many related manganese oxide minerals contain OH$^-$ groups [8]. First-principles calculations can help clarify these structural issues.

We started with K$_2$X$_2$Mn$_{24}$O$_{48}$ structures, where X refers to OH$^-$, H$_2$O or H$_3$O$^+$. We tested all the possible sites for these four adsorbates in the (2 × 2) tunnels using total energy minimization calculations. Placing one K and one OH$^-$/H$_2$O/H$_3$O$^+$ in each of the two empty (2 × 2) tunnels are energetically preferred. Fig. 6 shows the ground-state structure of $K_2X_2Mn_{24}O_{48}$ (X refers to $OH^-/H_2O/H_3O^+$). Each cell, $Mn_{24}O_{48}$, contains two $K^+$ and $OH^-/H_2O/H_3O^+$ in the (2 × 2) tunnels (Figs. 6(a), (c) & (e)). The axes of the tunnels are the favorable sites for the adsorbates. In the equilibrium state, the O atom of $OH^-/H_2O/H_3O^+$ is positioned closer to the $K^+$ ion than the H atom(s) of $OH^-/H_2O/H_3O^+$ (Figs. 6(b), (d) & (f)). Table 1 lists the optimum distance between K and the O atom of $OH^-/H_2O/H_3O^+$. $OH^-$ has the closest distance to $K^+$, compared to $H_2O$ and $H_3O^+$. As seen in Fig. 6(b), $OH^-$ bonds with $K^+$, and forms a compound of KOH. The K-OH bond length is 2.51 Å. Due to electronic charge repulsion, $K^+$ to $H_3O^+$ has the longest K-O distance, 3.46 Å. Experimental probes that can measure the K-O distance inside the tunnels should therefore be able to determine whether O is present in the form of $H_3O^+$ or not.

The best agreement with the experimental lattice parameters occurs for the model with $H_2O$ in the columns, for which $a$ = 9.771, $b/3$ = 2.846, $c$ = 9.762, and $\beta$ = 90.022°. Replacing $H_2O$ with either $OH^-$ or $H_3O^+$ decreases the predicted cell volume. The decrease in volume for the $H_3O^+$ case may be related to a short O-H bond formed by one H in each $H_3O^+$ with an O in the $MnO_2$ framework (Figure 6(e)).

K-doped α-$MnO_2$-related materials have a potential application in carbon capture and storage [6]. During the synthesis of these materials, $H_2O$ molecules are found in the (2 × 2) tunnels of α-$MnO_2$. After heating the sample at 150 $^0$C, $H_2O$ can be removed from the sample, but K remains [6]. Our first-principles calculations indicate that the binding energy of $H_2O$ in the α-$MnO_2$ is 0.39 eV, smaller than that of K by 4 eV. Such a low binding energy implies a weak interaction of $H_2O$ with K-doped α-$MnO_2$, and that therefore it should be relatively easy to remove $H_2O$ from the α-$MnO_2$ and related materials. Completely removing $H_2O$ content in the sample increases the adsorption uptake of $CO_2$ [6].

| Adsorbate X | K-O distance (Å) |
| --- | --- |
| $OH^-$ | 2.51 |
| $H_2O$ | 2.67 |
| $H_3O^+$ | 3.46 |

Table 1 Optimum distance between K and $OH^-/H_2O/H_3O^+$

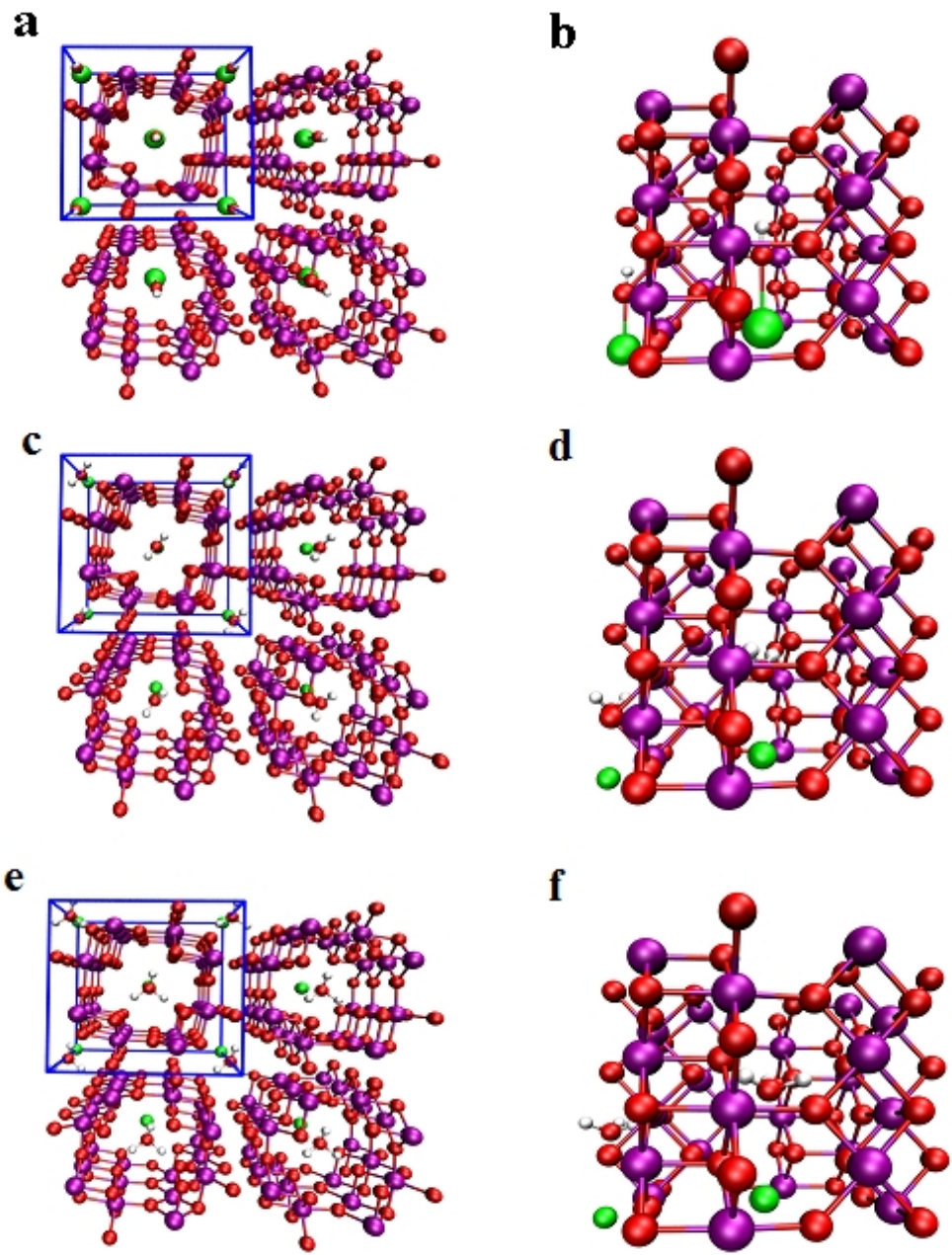

Fig. 6. $K_2X_2Mn_{24}O_{48}$, where X represents $OH^-/H_2O/H_3O^+$, K in green, Mn in purple, O in red and H in white: (a), (c) & (e) (2×2) supercell down tunnel axis; (b) (d) & (f) side view of a single cell.

4. **Conclusions**

First-principles density functional theory calculations were used to investigate α-$MnO_2$, a structure containing a framework of corner and edge sharing $MnO_6$ octahedra with tunnels in between. The calculated 1.3 eV band gap agrees with experiment, demonstrating the predictive power of DFT+U+J for manganese dioxide. The predictions of short-range antiferromagnetism, an indirect bandgap , and optimal K-O distances for tunnels containing both $K^+$ and $OH^-$, $H_2O$, or $H_3O^+$ show the variety of phenomena that occur in this system, and provide quantitative predictions for comparison with future experiments. Our results provide a benchmark for further computational studies of the technologically important family of manganese oxide materials.

**Acknowledgments**

We thank W. Wong-Ng and L. Espinal for helpful discussions and W. Wong-Ng for providing us with preliminary crystallographic data on α-$MnO_2$.